\documentclass[
    prl,
    superscriptaddress,
    amsmath,amssymb,amsfonts,
    reprint,
    aps,
    floatfix
]{revtex4-2}

\pdfoutput=1 
\newcommand{\papertitle}{Gearbox cavities: Ultrahigh wave confinement via broken rotational symmetry}

\usepackage{graphicx}
\usepackage{bm} 
\usepackage{physics}
\usepackage{amsthm}
\usepackage{amsmath}
\usepackage{xcolor}
\usepackage{textcomp}
\usepackage{outlines}
\usepackage{mathtools}
\usepackage{soul}
\usepackage{xr-hyper}
\usepackage[colorlinks=true, allcolors=black]{hyperref}
\usepackage[toc,page]{appendix}
\usepackage{cleveref}
\usepackage{multirow}

\renewcommand{\Re}{\operatorname{Re}}
\renewcommand{\Im}{\operatorname{Im}}

\newcommand{\Cv}[1]{C_{#1\mathrm{v}}}
\newcommand{\Cinfv}{C_{\infty\mathrm{v}}}

\renewcommand{\eqref}[1]{Eq.~(\ref{#1})}

\newcommand{\REVISION}[1]{{\color{red} #1}}
\DeclareUnicodeCharacter{202F}{\REVISION{There is a 202F unicode character here}}

\begin{document}

\preprint{APS/123-QED}

\title{\papertitle}

\author{Alessio Amaolo}
\affiliation{Department of Chemistry, Princeton University, Princeton, New Jersey 08544, USA}
\author{Sean Molesky}
\affiliation{Department of Engineering Physics, Polytechnique Montr\'{e}al, Montr\'{e}al, Qu\'{e}bec H3T 1J4, CAN}
\author{Alejandro W. Rodriguez}
\affiliation{Department of Electrical and Computer Engineering, Princeton University, Princeton, New Jersey 08544, USA}
\author{George Shaker}
\affiliation{Department of Mathematics, Massachusetts Institute of Technology, Cambridge, Massachusetts 02139, USA} 
\author{Steven G. Johnson}
\affiliation{Department of Mathematics, Massachusetts Institute of Technology, Cambridge, Massachusetts 02139, USA} 
\author{Pengning Chao}
\email{pchao827@mit.edu, pchao@maxwell-labs.com}
\altaffiliation{Present address: Maxwell Labs, St. Paul, MN, USA}
\affiliation{Department of Mathematics, Massachusetts Institute of Technology, Cambridge, Massachusetts 02139, USA}

\date{\today}
\begin{abstract}
Cavity enhancement of emission from a point source typically relies on structures that preserve its cylindrical symmetry, such as circular Bragg gratings.
Combining inverse design with fundamental limits, we show that breaking this symmetry is necessary to maximize the local density of states in lossless two-dimensional cavities.
Symmetry-broken designs exceed limits over all cylindrically symmetric structures by up to eight orders of magnitude, acting as ``gearboxes'' that convert the source's emission into high-angular-momentum waves trapped behind a centrifugal barrier.
\end{abstract}

\maketitle

\pagebreak
\begin{figure}
    \centering
    \includegraphics[width=0.95\linewidth]{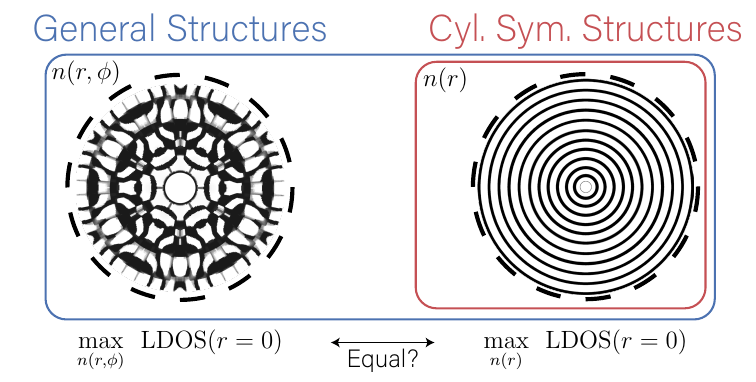}
    \caption{\textbf{Is a cylindrically symmetric cavity optimal?}
    The LDOS of a source at the center of a design region (dashed circle) is maximized either over all structures \(n(r, \phi)\) or over cylindrically symmetric structures \(n(r)\) alone, illustrated by a gearbox and a radial Bragg reflector, respectively.
    We show, for the regions and materials studied here, that general structures have provably larger maximum LDOS than their cylindrically symmetric counterparts.
    }
    \label{fig:schematic}
\end{figure}

Wave confinement in resonant cavities enhances wave--matter interactions and provides frequency selectivity, supporting applications in optics~\cite{vahala_optical_2003}, acoustics~\cite{ma_acoustic_2016}, and quantum corrals~\cite{fiete_colloquium_2003}.
Substantial effort has been devoted to optimal design of cavities for a given volume and set of materials (such as transparent dielectrics in optics), both by large-scale computational design (yielding lower bounds on attainable performance)~\cite{joannopoulos_photonic_2008, chen_validation_2024} and by recent progress on theoretical performance bounds over any possible structure~\cite{chao_physical_2022}.
A natural objective is the local density of states (LDOS)~\cite{novotny_principles_2012}, proportional to the power emitted by a point source (a dipole antenna in optics) of fixed amplitude and orientation at a given position and frequency.
When a single resonance at some complex frequency \(\tilde{\omega}_0\) dominates, the peak LDOS scales as \(Q/V\), where the quality factor \(Q = \Re \tilde{\omega}_0 / 2 \Im \tilde{\omega}_0\) characterizes the resonance lifetime and the effective mode volume \(V\) describes field concentration at the source~\cite{novotny_principles_2012}.

Symmetry plays a central role in wave confinement: breaking continuous translational symmetry down to a discrete one opens bandgaps, and breaking the remaining discrete symmetry with a defect localizes modes within them~\cite{joannopoulos_photonic_2008}.
For radial waves emitted by a point source, the standard solutions break only radial translational symmetry and preserve cylindrical symmetry about the source~\cite{joannopoulos_photonic_2008, jebali_grating_2007}.
The concentric layers of a bullseye also form the cross section of a Bragg fiber, which guides waves along its axis~\cite{joannopoulos_photonic_2008}, while concentric spherical shells form a Bragg onion~\cite{xu_highquality_2003}.
Each additional layer of a radial Bragg reflector suppresses outgoing radiation by a constant factor, so the LDOS grows exponentially with device size~\cite{jebali_grating_2007}.
Numerical optimization of the LDOS of an out-of-plane (TM) dipole, started from vacuum or from a photonic-crystal cavity, likewise often converges to bullseye-like concentric rings~\cite{liang_formulation_2013}.
Three-dimensional topology optimization has similarly produced nearly axisymmetric designs, motivating Ref.~\cite{christiansen_OrdersMagnitudeReduction_2026} to constrain searches to axisymmetric structures surrounded by Bragg-like shells.
Preserving cylindrical symmetry is also intuitive: a point source couples only to the cylindrically symmetric (\(m = 0\)) component of the field, so azimuthal scattering diverts power from the only channel that reaches the source and is often modeled as a parasitic loss~\cite{igarashi_Simicroringresonatorsidewall_2023, chen_CoexistenceRadialGuidedMode_2024}.
Still, this structure is not guaranteed to be optimal. LDOS maximization is a non-convex problem of the geometry (were it convex, averaging an optimal structure over all rotations about the source would yield an equally performing cylindrically symmetric structure), so even a problem invariant under every rotation and reflection about the source can have optima that spontaneously break this symmetry.
Similar symmetry breaking has been observed for in-plane dipoles, where local optimization of the polarization-averaged LDOS consistently favored one polarization at the expense of the other~\cite{liang_formulation_2013}.
These considerations raise the question: \textbf{is a cylindrically symmetric cavity optimal for enhancing the LDOS?}

In this paper, we prove that cylindrically symmetric cavities are not optimal (Fig.~\ref{fig:schematic}).
Specifically, we compute fundamental limits~\cite{chao_maximum_2022, shim_fundamental_2019, molesky_global_2020} over all cylindrically symmetric structures and exhibit structures that break the symmetry and exceed them; such a \textit{certificate of suboptimality} rules out the entire symmetric class and requires neither a tight limit nor an optimal design.
For example, for a source surrounded by a thin annulus of width \(0.8\lambda\), symmetry-broken designs reach an LDOS enhancement of \(1.5 \times 10^9\) at an outer radius of \(5.4\lambda\) (lossless material, single frequency), roughly eight orders of magnitude above the limit for cylindrically symmetric annuli.
We term the resulting structures \textit{gearbox cavities}.
A thin annulus with \(N\)-fold rotational symmetry acts as a \textit{gear}, converting the cylindrically symmetric wave emitted by the source into waves whose angular momentum is a multiple of \(N\); cascaded gears form a \textit{gearbox}, which raises the angular momentum in one step or through successively higher orders, like a mechanical gear train.
Intuitively, a wave of high angular momentum is held near the cavity by a centrifugal barrier and escapes only by tunneling through it.
Related azimuthally modulated structures include gear-shaped microdisk lasers~\cite{fujita_microgear_2002}, microrings with sidewall or azimuthal gratings~\cite{igarashi_Simicroringresonatorsidewall_2023, chen_CoexistenceRadialGuidedMode_2024}, and angular-grating emitters that couple high-angular-momentum modes into radiating channels~\cite{lu_highly-twisted_2023}. 
In these systems, the modulation acts on a whispering-gallery mode that already carries high angular momentum. 
By contrast, the gearbox converts a source's emission into confined high-angular-momentum waves, and our certificate establishes that breaking cylindrical symmetry is necessary to maximize the LDOS.

\paragraph{Formulation.} \label{sec:formulation}
We work in two dimensions for numerical and conceptual simplicity; the symmetry arguments below have direct three-dimensional analogs.
A time-harmonic scalar field \(\psi(\mathbf r)\) in an inhomogeneous medium obeys the Helmholtz equation, and maximizing the LDOS at a point \(\bm r'\) over all structures in a design region \(D \subset \mathbb R^2\) can be written as
\begin{equation}
\begin{aligned} \label{eq:ldos_opt}
    \max_{n(\mathbf{r})} \quad & \rho(\bm r') = \operatorname{Im} G(\bm r', \bm r') \\
    \textrm{s.t.} \quad & \nabla^2 G(\mathbf r, \bm r') + k_0^2\, n^2(\mathbf r)\, G(\mathbf r, \bm r') = -\delta(\mathbf r - \bm r') \\
    & n^2(\mathbf r) \in \{1,\, 1+\chi\} \quad \forall \mathbf r \in D
\end{aligned}
\end{equation}
where \(G(\mathbf r, \bm r')\) is the outgoing Green's function, \(k_0 = 2\pi/\lambda\) is the background wavenumber, \(n(\mathbf r)\) is the local refractive index (\(n = 1\) outside \(D\)), \(\chi > 0\) is the material contrast available within \(D\), and \(\rho_0\) is the value of \(\rho\) in the homogeneous background (\(n = 1\) everywhere).
For a \(z\)-polarized (TM) electromagnetic field, Maxwell's equations reduce exactly to \eqref{eq:ldos_opt} with \(\psi = E_z\), \(n^2\) the relative permittivity, and \(\chi\) the electric susceptibility, so that \(\rho/\rho_0\) is the Purcell enhancement.

We place the source at the origin of a circular design region, \(\bm r' = 0\), and expand the field in azimuthal harmonics \(\psi(r, \phi) = \sum_{m} c_m(r)\, e^{i m \phi}\), with \((r, \phi)\) polar coordinates centered on the source and \(c_m(r)\) the radial profile of the harmonic with angular momentum \(m\); only the \(m = 0\) harmonic is nonzero at the origin.
Every harmonic with \(m \neq 0\) is suppressed by the centrifugal barrier for \(r < r_m = \abs{m}/k_0\) and propagates as a traveling wave in the background beyond it.
In cylindrically symmetric (\(\Cinfv\)) cavities, angular momentum is conserved and \eqref{eq:ldos_opt} reduces to a one-dimensional radial problem.
Breaking cylindrical symmetry couples the harmonics: if \(n^2(r, \phi)\) has \(N\)-fold discrete rotational symmetry (\(C_N\)), the \(m = 0\) channel couples to harmonics \(m \in N\mathbb Z\) (adding a mirror plane to form \(\Cv{N}\) sets \(c_{-m} = c_m\)).
We obtain designs by computational optimization~\cite{molesky_inverse_2018, christiansen_OrdersMagnitudeReduction_2026, christiansen_inverse_2021} and compare them against fundamental limits~\cite{chao_maximum_2022, shim_fundamental_2019, molesky_global_2020} for each symmetry class, computed for lossless materials at zero bandwidth~\cite{amaolo_lossless_2026} to isolate geometric effects from absorption and detuning (see Supplemental Material (SM)~\cite{SM}).

\section{Results}

\paragraph{Optimality of Symmetry Breaking.}

\begin{figure*}
    \centering
    \includegraphics[width=\linewidth]{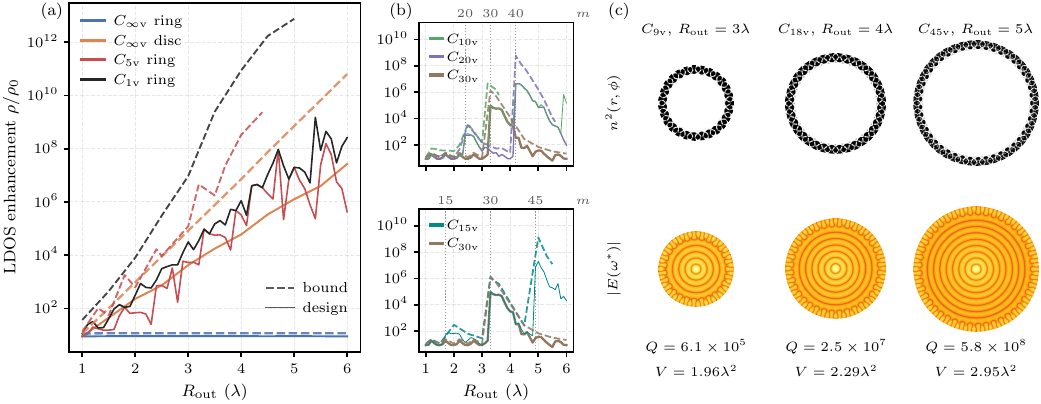}
    \caption{\textbf{Symmetry breaking is necessary for optimal Purcell enhancement.}
        All results are for a \(z\)-polarized (TM) dipole at the origin of a two-dimensional design region with lossless susceptibility \(\chi=2\).
        Designs are evaluated at their resonances \(\omega^*\), within \(1\%\) of \(\omega = 2\pi/\lambda\).
        Dashed lines: limits over all structures in each symmetry class; solid lines: inverse designs.
        Limit curves terminate where the solver could not find an initial dual-feasible point, a numerical failure since such points exist~\cite{amaolo_lossless_2026} and the LDOS is therefore finite~\cite{hsu_Boundstatescontinuum_2016}.
        \textbf{(a)} LDOS enhancement \(\rho/\rho_0\) versus outer radius \(R_\textrm{out}\) for \textit{rings} of fixed width \(w = R_\textrm{out}-R_\textrm{in} = 0.8\lambda\), with cylindrically symmetric (\(\Cinfv\)) \textit{discs} of fixed inner radius \(R_\textrm{in}=0.2\lambda\) for comparison.
        Each \(\Cv{N}\) design curve shows the best ring among all tested orders divisible by \(N\): \(5, 10, \ldots, 45\) for \(\Cv{5}\) and \(3\) to \(55\) for \(\Cv{1}\).
        Since every \(C_{\ell Nv}\) structure (\(\ell \in \mathbb{Z}^+\)) also has \(C_{Nv}\) symmetry, the \(C_{\ell Nv}\) limit cannot exceed the \(C_{Nv}\) limit.
        \(\Cinfv\) rings saturate at \(\rho/\rho_0 \approx 10\) (designs and limit), whereas symmetry-broken rings grow exponentially.
        \textbf{(b)} Limits and designs versus \(R_\textrm{out}\) for symmetry orders that are multiples of \(10\) (top) and \(15\) (bottom).
        A \(\Cv{N}\) ring couples the \(m=0\) channel only to \(m \in N\mathbb{Z}\), so both limits and designs exhibit windows of enhanced LDOS where such a mode is supported. Dotted lines mark the smallest \(R_\textrm{out}\) at which the best design has a new dominant harmonic \(m\), the largest \(\abs{c_m}^2\) of its field just outside the ring.
        Every peak of \(\Cv{\ell N}\) is also a peak of \(\Cv{N}\) at the same \(m\).
        \textbf{(c)} Best structures \(n^2(r, \phi)\) and driven fields \(\abs{E(\omega^*)}\) (log scale) at \(R_\textrm{out} = 3\lambda\), \(4\lambda\), and \(5\lambda\).
    }
    \label{fig:annulus}
\end{figure*}

To drive intuition on why optimal structures break the radial symmetry of \eqref{eq:ldos_opt}, we first consider annular design regions of fixed width \(w = 0.8\lambda\) and varying outer radius \(R_\textrm{out}\) (Fig.~\ref{fig:annulus}).
Cylindrically symmetric rings show no improvement with \(R_\textrm{out}\). 
As expected, a single ring reflects the outgoing \(m = 0\) wave with a fixed reflectivity, independent of its radius.
Similarly, \(\Cinfv\) discs, formed by cascaded rings, scale exponentially in both designs and limits.
In stark contrast, symmetry-broken \textit{rings} improve exponentially with \(R_\textrm{out}\), exceeding the \(\Cinfv\) ring limit by a factor of \(1.3\times10^{8}\) at \(R_\textrm{out} = 5.4\lambda\), and outperform \(\Cinfv\) disc designs (\(R_\textrm{in} = 0.2\lambda\)) of equal \(R_\textrm{out}\) at nearly every radius, though not the \(\Cinfv\) disc limit.
The same certificate of suboptimality holds for lossy rings (\(\chi = 5 + 10^{-6}i\)): symmetry-broken designs exceed the \(\Cinfv\) ring limit at every \(R_\textrm{out}\), and for \(R_\textrm{out} \leq 1.5\lambda\) they exceed the \(\Cinfv\) limit for the disc containing them (see SM~\cite{SM}).
The optimized rings are gears with periodic teeth along the azimuth (Fig.~\ref{fig:annulus}(c), hence the name); their LDOS grows through the exponential scaling of their \(Q\)-factors, over ten orders of magnitude, while the mode volume varies between \(V \approx 0.3\) and \(50\lambda^2\).
The corresponding field profiles suggest the mechanism is a standing \(m = 0\) wave in the center while the ring carries high-\(m\) components resembling whispering-gallery modes (WGMs).
As \(R_\textrm{out} \to \infty\), the gear becomes locally a linear waveguide grating under a normally incident plane wave, which it can reflect perfectly by critically coupling it to a guided mode~\cite{quaranta_recent_2018}.
The gear analogously reflects the \(m = 0\) wave by coupling it to WGMs, with a reflectivity capped by curvature in the same way as the \(Q\) of a smooth ring~\cite{snyder_optical_1983}, so that both improve exponentially as the curvature decreases.

As explained above, a \(\Cv{N}\) gear couples \(m = 0\) only to harmonics \(m \in N\mathbb{Z}\).
The phase of harmonic \(m\) advances by \(2\pi m\) around the ring, so a wave in the vacuum outside can be phase matched to it, and carry its energy away, only if \(m\) vacuum wavelengths fit around the outer circumference, \(m \leq k_0 R_\textrm{out}\).
A ring therefore traps a whispering-gallery mode of order \(m\), as in total internal reflection, only if it is phase matched to waves in the material, whose wavelength is \(\lambda/\sqrt{1+\chi} = \lambda/\sqrt{3}\), but not to those in vacuum, \(k_0 R_\textrm{out} \lesssim m \lesssim \sqrt{3}\, k_0 R_\textrm{out}\); equivalently, a confined wave must propagate in the material but be evanescent in vacuum.
Consequently, a gear of fixed symmetry should be effective only at particular \(R_\textrm{out}\).
Correspondingly, each \(\Cv{N}\) class in Fig.~\ref{fig:annulus}(b) shows sharp peaks in both inverse designs and limits. 
Because the orders allowed by \(\Cv{\ell N}\), for any positive integer \(\ell\), are a subset of those allowed by \(\Cv{N}\), every peak of \(\Cv{\ell N}\) is also a peak of \(\Cv{N}\), at the same \(m\).
The dotted lines mark the smallest \(R_\textrm{out}\) at which a new harmonic \(m\) has the largest azimuthal Fourier amplitude \(\abs{c_m}^2\) of the design's field just outside the ring (\(0.05\lambda\) beyond \(R_\textrm{out}\)).
Every design curve rises at this radius and peaks within \(0.1 \lambda\) of it, by more than five orders of magnitude between neighboring radii, where the gear begins converting \(m = 0\) into the next allowed harmonic; the more coarsely sampled limits rise in the same windows.
For \textit{every} class from \(\Cv{5}\) to \(\Cv{45}\), this conversion sets in at \(1.3\, k_0 R_\textrm{out} \lesssim m \lesssim 1.5\, k_0 R_\textrm{out}\), where the dominant harmonic is evanescent outside the ring.
Because each class performs well only near its own peaks, the \(\Cv{1}\) ring designs, calculated via a maximum of the best \(\Cv{N}\) designs for \(N \in [3,55]\) at each \(R_\textrm{out}\) form an envelope over the peaks of all classes.

\paragraph{Gearbox Cavities.}

\begin{figure*}
    \centering
    \includegraphics[width=\linewidth]{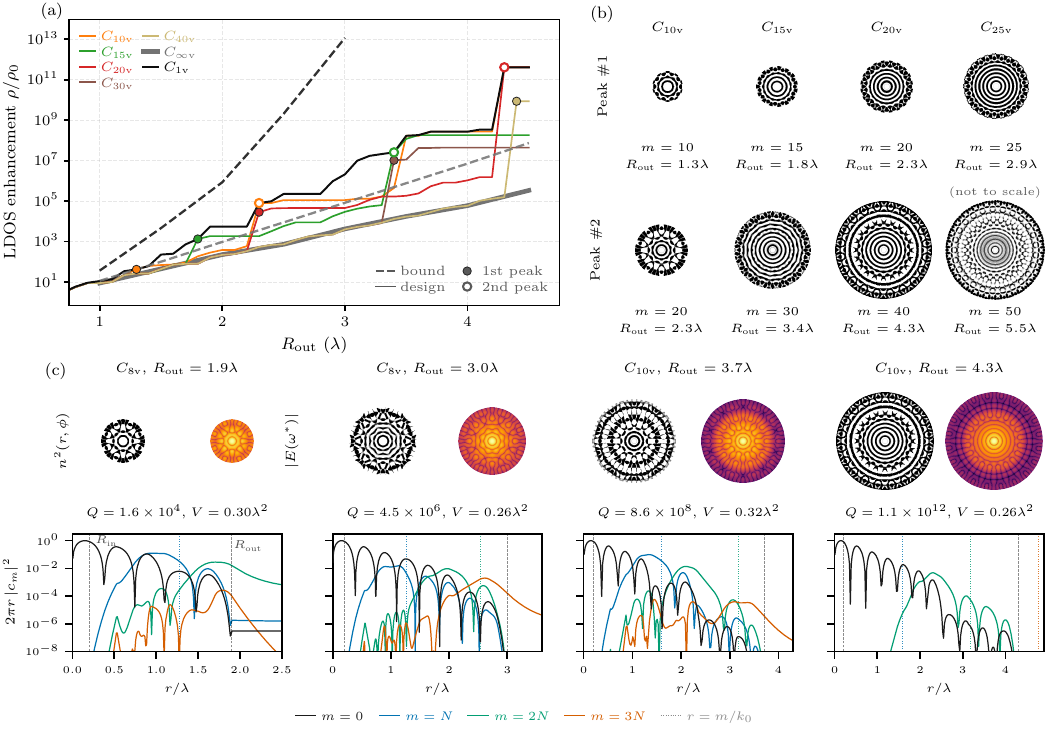}
    \caption{\textbf{Gearbox discs exceed the cylindrically symmetric disc limit.}
    Same source, material, and frequency as Fig.~\ref{fig:annulus}, for \textit{discs} of fixed inner
    radius \(R_\textrm{in}=0.2\lambda\).
    Disc design regions are nested (a disc can leave its outer shells empty), so every design curve is a
    running maximum in \(R_\textrm{out}\).
    \textbf{(a)} \(\rho/\rho_0\) as a function of \(R_\textrm{out}\).
    Gray: \(\Cinfv\) designs (solid) and limit (dashed).
    Colored: designs of a single order \(N = 10, 15, 20, 30, 40\).
    Black: the best disc over every order computed, \(N = 3\) to \(45\) (solid), and the limit over all
    mirror-symmetric (\(\Cv{1}\)) discs (dashed), which ends at \(R_\textrm{out} = 3\lambda\) for the same numerical reason given in Fig.~\ref{fig:annulus}.
    Filled and hollow circles mark the first and second radii at which a new harmonic dominates the field just outside the disc, as in Fig.~\ref{fig:annulus}(b); on these running maxima they coincide with steps in performance, which we call peaks.
    Symmetry-broken discs exceed the \(\Cinfv\) limit at every computed \(R_\textrm{out} \geq 1.5\lambda\), by a
    factor of \(5.5\times10^{3}\) at \(4.5\lambda\).
    \textbf{(b)} Structures at the first (Peak \#1) and second (Peak \#2) peaks of \(\Cv{10}\) through
    \(\Cv{25}\), labeled with channel \(m\) and \(R_\textrm{out}\).
    All structures in (b) and (c) share one radial scale, except the \(R_\textrm{out} = 5.5\lambda\) structure (not to scale).
    \#1 designs are the smallest that support the \(m=N\) whispering-gallery mode, so they are radial Bragg
    reflectors capped by a single gear, which converts \(m=0\) into \(N\); \#2 designs add a second gear, converting
    \(N\) into \(2N\) when the gearbox is large enough to support both modes.
    \textbf{(c)} Best discs at \(R_\textrm{out} = 1.9\lambda\), \(3\lambda\), \(3.7\lambda\), and \(4.3\lambda\) along with structures \(n^2(r, \phi)\) and driven fields \(\abs{E(\omega^*)}\) (log scale).
    Below, the azimuthal content of the resonant mode, \(2\pi r\abs{c_m}^2\), summed over \(\pm m\) and
    normalized to the peak of the total.
    Dashed lines mark \(R_\textrm{in}\) and \(R_\textrm{out}\); dotted lines mark the vacuum turning points
    \(r_m = m/k_0\), beyond which channel \(m\) propagates.
    Power leaves \(m=0\) within the inner rings, passes through successively higher multiples of \(N\), and
    exits in the highest one populated (\(m = 16\), \(24\), \(30\), and \(40\)), whose turning point lies
    beyond \(R_\textrm{out}\).
    The \(R_\textrm{out} = 4.3\lambda\) disc, though optimized as \(\Cv{10}\), converged to a \(\Cv{20}\) structure, so it carries no \(m = 10\), and its exit channel \(m = 40\) (\(r_{40} \approx 6.4\lambda\)) lies beyond the radii shown.}
    \label{fig:disc}
\end{figure*}

We next consider discs of fixed inner radius \(R_\textrm{in} = 0.2\lambda\) (Fig.~\ref{fig:disc}).
A disc can hold many concentric rings, each reflecting the outgoing wave. 
As expected, the resulting \(\Cinfv\) structures form Bragg reflectors with exponentially growing LDOS in both designs and limits.
Symmetry-broken discs exceed the \(\Cinfv\) disc limit at every computed \(R_\textrm{out} \geq 1.5\lambda\), by a factor of \(5.5\times10^{3}\) at \(R_\textrm{out} = 4.5\lambda\), again certifying the suboptimality of cylindrically symmetric structures.
The best disc improves little beyond \(R_\textrm{out} = 4.3\lambda\), where it reaches \(Q \sim 10^{12}\): at such \(Q\), double precision fixes the resonance frequency only to about half a linewidth, so the LDOS that the optimizer maximizes at that frequency becomes unreliable.
Limits over mirror-symmetric (\(\Cv{1}\)) structures, containing every design shown, lie well above the best rings and discs as \(R_\textrm{out}\) grows; whether this gap reflects loose limits or undiscovered designs remains open.

Again, a \(\Cv{N}\) disc couples \(m = 0\) only to \(m \in N\mathbb{Z}\) and supports a whispering-gallery mode of order \(m\) only within the same window, \(k_0 R_\textrm{out} \lesssim m \lesssim \sqrt{3}\, k_0 R_\textrm{out}\), so each class should improve abruptly whenever \(R_\textrm{out}\) admits a new multiple of \(N\).
Indeed, each \(\Cv{N}\) curve in Fig.~\ref{fig:disc}(a) steps up where a new harmonic begins to dominate the field just outside the disc, occurring at \(1.2\, k_0 R_\textrm{out} \lesssim m \lesssim 1.5\, k_0 R_\textrm{out}\) for every marked peak.
Below its first peak a \(\Cv{N}\) disc is a radial Bragg reflector; the first peak caps it with a single gear converting \(m = 0\) into \(N\), and the second adds a gear converting \(N\) into \(2N\), forming a gearbox (Fig.~\ref{fig:disc}(b)).
The exception is \(\Cv{5}\): its turning point \(r_5 \approx 0.8\lambda\) lies inside every disc shown, so \(m = 5\) cannot be trapped, and its first peak is instead effectively a \(\Cv{10}\) gear converting \(m = 0\) directly into \(10\), matching the \(\Cv{10}\) disc at that radius.
A \(\Cv{2N}\) disc admits no harmonic between \(0\) and \(2N\) and so must convert in a single step.
Where the second peak of \(\Cv{N}\) and the first of \(\Cv{2N}\) reach the same channel \(m = 2N\) at nearly the same \(R_\textrm{out}\), as for \(N = 10\), \(15\), and \(20\), the two-gear design performs better, by up to a factor of \(47\).
Every \(\Cv{2N}\) disc is also a \(\Cv{N}\) disc but optimizations converged on the two-step gearbox, suggesting that, at least for these orders, raising the angular momentum in two steps is more effective than in one.
The azimuthal spectra of the best discs (Fig.~\ref{fig:disc}(c)) show the cascade directly: power leaves \(m = 0\) within the Bragg core, passes through successively higher multiples of \(N\), and exits in the highest channel populated, whose turning point lies beyond \(R_\textrm{out}\), so that it escapes only by tunneling through the centrifugal barrier and the radiating wave is converted into a trapped wave.
As for rings, the LDOS grows through \(Q\) at nearly constant mode volume: \(Q\) reaches \(1.1\times10^{12}\) at \(R_\textrm{out} = 4.3\lambda\) while \(V\) stays near \(0.3\lambda^2\).

\section{Conclusions} \label{sec:conclusions}
We have shown that breaking cylindrical symmetry is necessary to maximize the LDOS of a point source for the regions and materials studied here: inverse-designed gearbox cavities, which convert the source's emission into high-angular-momentum waves trapped behind a centrifugal barrier, exceed limits over all cylindrically symmetric structures by up to eight orders of magnitude.
More broadly, certificates of suboptimality use fundamental limits not as benchmarks for individual designs but to decide which classes of structure can contain the optimum, a question that arises whenever symmetry, fabrication, or other constraints restrict the design space.
Looking ahead, we anticipate that gearboxes can be fine-tuned for photonic slabs of finite thickness, that chiral (\(\mathrm{C}_N\)) gearboxes, which can populate \(+m\) and \(-m\) unequally, may serve as compact sources of orbital angular momentum, and that the same mechanism carries over to other scalar waves such as sound in acoustic resonators.
An open question is whether gears, whose confinement grows with radius, can be cascaded so that their confinement factors multiply, as the constant factors of Bragg layers do, making the LDOS grow superexponentially with device size.

\begin{acknowledgments}
\paragraph{Acknowledgments.} A substantial portion of the simulations presented in this article were performed on computational resources managed and supported by Princeton Research Computing, a consortium of groups including the Princeton Institute for Computational Science and Engineering (PICSciE) and Research Computing at Princeton University.
The authors also acknowledge the MIT SuperCloud and Lincoln Laboratory Supercomputing Center for providing HPC resources that have contributed to the research results reported within this paper.

\paragraph{Funding.} S.M. acknowledges financial support from NSERC under Discovery Grant RGPIN-2023-05818, as well as additional benefits provided by affiliations to the Regroupement Québécois sur les Matériaux de Pointe, \url{https://doi.org/10.69777/309032}, the IVADO Research Consortium, and the Lassonde Deeptech Institute. 
S.G.J. is grateful for support from the Simons Foundation.
A.A. and A.W.R. acknowledge the support by a Princeton SEAS Innovation Grant.
Research was sponsored by the Army Research Office and was accomplished under Cooperative Agreement Number W911NF-26-2-A132. 
The views and conclusions contained in this document are those of the authors and should not be interpreted as representing the official policies, either expressed or implied, of the Army Research Office or the U.S. Government. The U.S. Government is authorized to reproduce and distribute reprints for Government purposes notwithstanding any copyright notation herein.
\end{acknowledgments}

\bibliographystyle{apsrev4-2}
\bibliography{refs}

\end{document}


\title{Supplemental Material for ``\papertitle''}

\author{Alessio Amaolo}
\affiliation{Department of Chemistry, Princeton University, Princeton, New Jersey 08544, USA}
\author{Sean Molesky}
\affiliation{Department of Engineering Physics, Polytechnique Montr\'{e}al, Montr\'{e}al, Qu\'{e}bec H3T 1J4, CAN}
\author{Alejandro W. Rodriguez}
\affiliation{Department of Electrical and Computer Engineering, Princeton University, Princeton, New Jersey 08544, USA}
\author{George Shaker}
\affiliation{Department of Mathematics, Massachusetts Institute of Technology, Cambridge, Massachusetts 02139, USA} 
\author{Steven G. Johnson}
\affiliation{Department of Mathematics, Massachusetts Institute of Technology, Cambridge, Massachusetts 02139, USA} 
\author{Pengning Chao}
\email{pchao827@mit.edu, pchao@maxwell-labs.com}
\altaffiliation{Present address: Maxwell Labs, St. Paul, MN, USA}
\affiliation{Department of Mathematics, Massachusetts Institute of Technology, Cambridge, Massachusetts 02139, USA}

\date{\today}

\maketitle

\section{Resonant Waveguide Gratings and Photonic Gearboxes} \label{asec:gratings_theory}

\begin{figure}
    \centering
    \includegraphics[width=\linewidth]{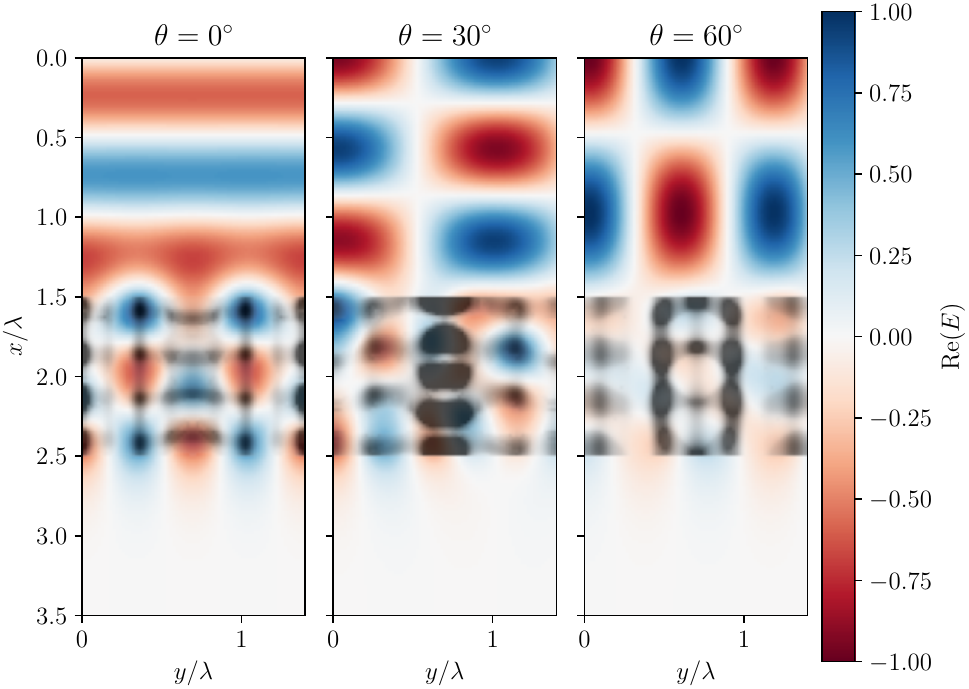}
    \caption{\textbf{Linear resonant waveguide gratings.}
    Linear resonant waveguide gratings that are inverse-designed to perfectly reflect planewaves at specific incident angles. The design region has size $1\lambda\times1.4\lambda$ and is periodic in the $y$ direction. Material used has relative permittivity $\epsilon_r=6$. The incident planewaves have wavevectors at angle $\theta = 0^\circ, 30^\circ, 60^\circ$ to the $x$ direction. The total field is plotted with a translucent overlay of the gratings on top. All designs have mirror symmetry with respect to the $x$ axis and have transmittance $<10^{-9}$ for the designed $\theta$. 
    }
    \label{fig:lrwg}
\end{figure}

In this section, we describe in more detail how gear cavities connect asymptotically to linear resonant waveguide gratings (RWGs) and present an argument for why \textit{super-exponential} scaling of confinement in gearbox cavities (iterated concentric gears) may be feasible. 

First, consider the linear RWG, which can be designed to perfectly reflect an incident planewave through precise coupling of the incident wave with the leaky modes in the grating, canceling out the transmitted wave through interference with radiation of the leaky modes~\cite{quaranta_recent_2018}. Note that this perfect reflection is not restricted to normal incidence; for example, Figure \ref{fig:lrwg} demonstrates some inverse-designed linear RWGs with perfect reflection of planewaves incoming at various angles. 

We can understand the improved performance of gear cavities as their radius $R$ increases by connecting to linear RWGs in the $R \rightarrow \infty$ limit through the relevant asymptotics of cylindrical waves. First, let us consider the single gear that reflects an outgoing order $m$ cylindrical wave $H_m^{(1)}(kr)e^{im\phi}$; for a fixed order $m$, the relevant asymptotic is~\cite{dlmf}
\begin{equation}
    H_m^{(1)}(kr)e^{im\phi} \sim \sqrt{\frac{2}{\pi kr}} \exp\left[i\Phi(r,\phi)\right]
    \label{eq:asymptotics_fixed_m}
\end{equation}
with the phase of the wave given by
\begin{equation}
    \Phi(r,\phi) = kr + m\phi - \frac{m\pi}{2} - \frac{\pi}{4}
    \label{eq:asymptotics_fixed_m_phase}
\end{equation}
as $r\rightarrow\infty$. The local wavevector component of the wave at a given point $(r,\phi)$ in polar coordinates is then
\begin{align}
    k_r &= \pdv{\Phi}{r} = k \\
    k_\phi &= \frac{1}{r}\pdv{\Phi}{\phi} = \frac{m}{r} \to 0. 
\end{align}
We see that for a fixed order $m$, at large $r$ the wave will locally resemble a planewave traveling in the radial direction (of course, for larger $m$ the onset of this asymptotic regime occurs at larger $r$). Thus, as we increase the design radius of a single gear, at any local section of the gear we asymptotically approach a linear RWG with a normally incident planewave for which perfect reflection is possible, explaining the increased confinement of the gear. 

Now let us consider the gearbox scenario, where we attempt to further boost confinement through iterated gears. Now, we have not just a single order $m$ wave but a whole series of cylindrical waves with progressively increasing order. The argument of the previous paragraph suggests that if we iterate gears designed to reflect locally normal incident waves, a cylindrical wave of any given order will eventually see ever increasing reflectivity as it travels outwards. Iteratively multiplying a series of increasing reflectivities is a path towards super-exponential confinement, provided phase matching can also be achieved. However, this argument is not yet complete: as can be observed in Figure 3 of the main text, as the design radius increases, the highest relevant $m$ order also increases linearly with it. The cylindrical wave asymptotics given in Eqs.~(\ref{eq:asymptotics_fixed_m}) and~(\ref{eq:asymptotics_fixed_m_phase}) for fixed $m$ are no longer valid in this scenario; we must instead use Debye's expansions~\cite{dlmf} for simultaneously large order and argument:
\begin{equation}
    H_m^{(1)}(m \sec\beta)e^{im\phi} \sim \sqrt{\frac{2}{\pi m \tan\beta}} \, \exp(i\Phi)
\end{equation}
with phase
\begin{equation}
    \Phi = m \tan\beta - m\beta - \frac{\pi}{4} + m\phi,
\end{equation}
where we have written the radial argument $kr = m\sec\beta$ with $0<\beta<\pi/2$. Again, we can take partial derivatives of $\Phi$ to obtain the local wavevector components (expressing $\beta$ as function of $r$)
\begin{align}
    k_r &= \pdv{\Phi}{r} = \frac{m\tan\beta}{r}= k\sin\beta, \\
    k_\phi &= \frac{1}{r}\pdv{\Phi}{\phi} = \frac{m}{r} = k \cos\beta. 
\end{align}
We see that the scenario where the wave order increases linearly with the gear radius asymptotes to a linear RWG with an \textit{oblique incidence} planewave coming at an angle $\pi/2-\beta$ relative to the normal. Linear RWGs can also be engineered to achieve perfect reflection  for oblique incidence, as illustrated in Figure \ref{fig:lrwg}, so even if we linearly increase the wave order $m$ as we increase the gear radius, we can still achieve increasing gear reflectivity. 

Taken together, the fixed $m$ order asymptotics and the Debye expansion suggest a potential recipe for achieving super-exponential scaling: take a linear RWG designed for perfect reflection of planewaves at both normal incidence and oblique incidence at a fixed angle $\pi/2 - \beta$. Bend this linear RWG into rings of various radii and concentrically layer these rings around each other at appropriate radius intervals, so the increase in relevant wave order $m$ conforms with the asymptotics $m \sim kr \cos\beta$. This results in increasing layer-by-layer reflectivity both for any cylindrical wave of given $m$ as well as the ``leading edge'' of cylindrical waves with linearly increasing $m$. If the phase matching between the rings can also be finetuned (perhaps through some further structural inverse design) so the layer-by-layer reflectivities contribute to overall confinement in some iteratively multiplying fashion, then super-exponential scaling would have been achieved. The feasibility of doing such phase matching remains an open question for future work, though we hope that the asymptotic arguments of this section (as well as the strikingly high performance of gearbox inverse designs in Figure 3 of the main text) will convince the reader that such super-exponential size scaling is not as obviously impossible as it may seem on first glance.

\section{Calculation of Fundamental Limits} \label{asec:limits}
\paragraph{General Methodology.}
To calculate general limits for photonic objectives over arbitrary structuring, we employ methods in~\cite{molesky_global_2020, chao_physical_2022, chao_blueprints_2025}.
Here we summarize these methods for completeness, but do not duplicate derivations.
The fundamental photonic optimization problem we treat is
\begin{equation} \begin{aligned} \label{eq:emopt}
    \max_{\varepsilon (\mathbf r)} \quad & f(\mathbf E) \\ 
    \textrm{s.t.} \quad & \mathbf \nabla \times \mathbf \nabla \times \mathbf E - \omega^2 \varepsilon(\mathbf r) \mathbf E = i\omega \mathbf J
\end{aligned} \end{equation}
with incident source \(\mathbf J\) at frequency \(\omega\) and \(f\) a quadratic function of the resulting electric field \(\mathbf E\).
This problem can be generally re-written as an optimization of the polarization degrees of freedom \(\mathbf P(\mathbf r) = \chi(\mathbf r) \mathbf E (\mathbf r)\) by following the procedure in Ref.~\cite{chao_physical_2022}:
\begin{subequations}\begin{align} \label{eq:QCQP}
    \max_{\mathbf P} \quad & f(\mathbf P) \\ 
    \textrm{s.t.} \quad & \int_{\Omega_j} \left[ \mathbf P^* \cdot \left( \chi^{-\dagger} - \mathbb G_0^\dagger \right) \mathbf P - \mathbf E_i^* \cdot \mathbf P\right] = 0 \label{subeq:poynting} 
\end{align} \end{subequations}
where we have overloaded the definition of \(f\), \(\mathbb G_0\) is the vacuum Green's function that acts on \(\mathbf P\) via convolution with the inverse of the Maxwell operator, \(\Omega_j\) indicates every point in the design region, \(\mathbf E_i\) is the incident field caused by \(\mathbf J\) in vacuum, and most notably \(\chi\) is constant instead of spatially dependent.
Notably, we have traded a single PDE constraint by infinitely many complex scalar integral constraints~\eqref{subeq:poynting}, each of which represent energy conservation (Poynting's theorem) at each point in space.
Still,~\eqref{eq:emopt} and~\eqref{eq:QCQP} can be shown to be equivalent.
When \(f\) is a quadratic function~\eqref{eq:QCQP} forms a quadratically constrained quadratic program (QCQP) which can be relaxed using Lagrange duality~\cite{molesky_mathbbt-operator_2019} or semidefinite programming~\cite{gertler_sparse_2023}.
The relaxed problems are convex, and thus admit relatively straightforward global solutions that represent global limits to \eqref{eq:emopt} over any possible structure \(\varepsilon(\mathbf r)\).
It is also possible to generalize this framework to many contexts, including efficiency metrics~\cite{angeris_efficiency_2023}, multiple materials~\cite{amaolo_can_heterostructures_2023}, and nonlinear phenomena~\cite{amaolo_raman_2024, mohajan_fundamental_2023, chao_dist_raman_2025}. 

\paragraph{Sparse Formulation.}
Computing these limits requires the vacuum Green's function restricted to the design region \(D\), \(\mathbb G_{0,DD}\), which is dense as a dipole at any point yields a field at all points.
Following Ref.~\cite{chao_blueprints_2025}, we instead work with the Maxwell operator \(\mathbb M = -\nabla^2 - k_0^2\) (TM polarization), which is sparse in any local basis such as the finite-difference cells used here and satisfies \(\mathbb G_0 = k_0^2 \mathbb M^{-1}\).
Partitioning \(\mathbb M\) into blocks over design region \(D\) and the background \(B\), block inversion gives
\begin{equation} \label{eq:GinvDD}
    \mathbb G_{0,DD}^{-1} = \frac{1}{k_0^2} \left( \mathbb M_{DD} - \mathbb M_{DB} \mathbb M_{BB}^{-1} \mathbb M_{BD} \right),
\end{equation}
where \(\mathbb M_{XY}\) couples fields in region \(Y\) to region \(X\).
The second term is computed by sparse linear solves \(\mathbb M_{BB} \mathbb X = \mathbb M_{BD}\) and is nonzero only on the boundary between \(B\) and \(D\) so it is sparse.
Changing the optimization variable from \(\mathbf P\) to the scattered field \(\mathbb G_{0,DD} \mathbf P\) then expresses the objective and every constraint~\eqref{subeq:poynting} through \(\mathbb G_{0,DD}^{-1}\), \(\chi\), and the cell projectors alone, all of which are sparse.
The two problems are equivalent but the dual function and its gradient can now be evaluated with sparse factorizations instead of dense linear algebra~\cite{chao_blueprints_2025}.

\paragraph{Limits in Polar Coordinates.}
Every design region in this work is centered on the source, so we discretize \(\mathbb M\) on a polar grid.
Each cell is an annular sector of area \(r_j \Delta r \Delta\phi\), and these areas weight every integral in~\eqref{subeq:poynting}.
Following Ref.~\cite{lai_note_2001}, the radial nodes are at \(r_j = (j + \tfrac12)\Delta r\), so that no unknown lies at the coordinate singularity \(r = 0\). 
A point source is therefore a uniform current over the innermost ring of cells, \(r < \Delta r\).
Outgoing waves are absorbed by a radial stretched-coordinate perfectly matched layer (PML)~\cite{shin_choice_2012}.
Note that a harmonic \(m \neq 0\) is evanescent inside its turning point \(r_m = |m|/k_0\) and escapes only by tunneling through the centrifugal barrier, so we begin the PML beyond \(r_m\) of the highest harmonic the structures populate.

The polar grid also lets us compute limits on a single slice of the structure.
Consider a structure invariant under the group \(C_N\) of rotations by \(2\pi l/N\) for \(l \in \mathbb Z\).
Since \(\mathbb G_0\) commutes with rotations and the source is rotationally symmetric, the polarization shares the symmetry of the structure.
Thus, we specify polarizations on a wedge with periodic boundary conditions on its two edges.
A mirror plane (\(\Cv{N}\)) further halves the wedge to \(0 \le \phi \le \pi/N\), with Neumann boundary conditions on its mirror plane.
It therefore suffices to assemble \(\mathbb M\) on the wedge with these boundary conditions to apply \eqref{eq:GinvDD}.
Because \(\chi\) is constant and \(\mathbb G_0\) commutes with the symmetry operations, the constraint~\eqref{subeq:poynting} on a wedge cell \(\Omega_j\) coincides with those on its rotated images, so one constraint per wedge cell suffices. 
The objective computed on the wedge is reduced by a factor \(N\) (\(2N\) with a mirror plane), which cancels in the LDOS enhancement \(\rho/\rho_0\).
The \(\Cinfv\) limit is the \(N \to \infty\) case, a single azimuthal cell spanning the full circle.

To numerically calculate the limits in this paper, we employ our open source library Dolphindes~\cite{dolphindes} to assemble \(\mathbb M\) on the wedge and the corresponding sparse QCQP, and solve its dual via an interior point method augmented with generalized constraint descent as described in Ref.~\cite{chao_blueprints_2025}.

\section{Inverse Design Methodology} \label{asec:inverse_design}
To discover high-performance structures, we employ traditional adjoint-based topology optimization~\cite{jensen_topology_2011} with the analytical LDOS gradients available for example in Refs.~\cite{chao_maximum_2022, liang_formulation_2013, shaker_eigenvalue_2026}.
We employ the symmetry-reduced polar discretization of Sec.~\ref{asec:limits}, which solves Maxwell's equations on a single irreducible wedge of a \(\Cv{N}\) structure (a single azimuthal cell for \(\Cinfv\)) and is the natural basis for constructing gearboxes with the same symmetry structure.
We refer to this as the \textit{unshifted} algorithm, for which we use the method of moving asymptotes~\cite{svanberg_class_2002, svanberg_methodmovingasymptotes_1987} as implemented in the optimization library NLopt~\cite{johnson_nlopt_2019}. 
To finetune the structures once a resonance is ``caught'' (i.e., the structure exhibits a single resonant mode), we perform the \textit{shifted} algorithm as described in Ref.~\cite{shaker_eigenvalue_2026}, which allows the source frequency to drift in exchange for significantly better conditioning and convergence. 
We run the shifted algorithm first without any constraints on \(\omega\); we then run it as the inner loop of the Augmented Lagrangian method (as in NLopt~\cite{johnson_nlopt_2019}) with \(\omega^*=\omega\) as a constraint, terminating when \(\omega^*\) is within \(1\%\) of \(\omega\).
Thus, in principle, structures could be many linewidths from true resonance.
We consider this an acceptable tradeoff for high-performance structures: sources may be tuned or structures may be re-scaled to match the source frequency.

We do not binarize the final designs, so the susceptibility of each pixel takes values between \(0\) and \(\chi\).
For TM polarization, \(E_z\) is parallel to every in-plane interface, so a gray pixel is the effective medium of a subwavelength binary checkerboard whose permittivity is the volume average of \(1\) and \(1+\chi\).
Accordingly, we compare them against limits over structures resolved at least as finely.
The \(\Cinfv\) limits, against which every certificate in the main text is stated, have no azimuthal grid and are computed at the radial resolution of the designs.
The \(\Cv{1}\) ring limits plateau in azimuthal resolution for \(R_\textrm{out} \leq 3.5\lambda\), whereas the \(\Cv{1}\) disc limits are converged only for \(R_\textrm{out} \leq 1.5\lambda\), and those at \(R_\textrm{out} = 2.5\lambda\) and \(3\lambda\) were computed at half the radial resolution of the designs.
No claim in the main text rests on these \(\Cv{1}\) limits.

\paragraph{Definition of \(Q\) and \(V\).}
An outgoing source-free solution of Maxwell's equations for the TM field \(\tilde E \equiv \tilde E_z\) satisfies
\begin{equation} \label{eq:qnm}
    \nabla^2 \tilde E + \tilde k^2\, \varepsilon\, \tilde E = 0, \qquad \tilde k = \tilde\omega/c = k^* + i k_i,
\end{equation}
where \(\varepsilon = n^2\) is the relative permittivity, \(\tilde\omega = \omega^* + i\omega_i\) is the complex eigenfrequency, \(k^* = \omega^*/c\), and \(k_i = \omega_i/c\).
Under the \(e^{-i\omega t}\) time convention used throughout, the mode decays in time, so \(\omega_i < 0\) and 
\begin{equation}
    Q = \omega^*/2\abs{\omega_i} = k^*/2\abs{k_i}.
\end{equation}
We use the mode volume convention
\begin{equation} \label{eq:mode_vol}
    V = \frac{\int \varepsilon\, \tilde E^2 \, d^2\mathbf r}{\varepsilon(0)\, \tilde E(0)^2},
\end{equation}
where the integral runs over the computational domain, \(\tilde E(0)\) is the average of \(\tilde E\) over the source cells \(r < \Delta r\) (Sec.~\ref{asec:limits}), and \(\varepsilon(0) = 1\) in the vacuum core.
The LDOS \(\rho = \Im G(0, 0)\) is set by the outgoing Green's function \(G(\mathbf r, \mathbf r')\), which obeys the driven counterpart of~\eqref{eq:qnm} at the real drive wavenumber \(k = \omega/c\),
\begin{equation} \label{eq:green}
    \nabla^2 G(\mathbf r, \mathbf r') + k^2\, \varepsilon(\mathbf r)\, G(\mathbf r, \mathbf r') = -\delta(\mathbf r - \mathbf r').
\end{equation}
Near \(\omega^*\), \(G\) is dominated by the single-mode term of its modal expansion~\cite{novotny_principles_2012},
\begin{equation} \label{eq:G_single_mode}
    G(\mathbf r, \mathbf r') \approx \frac{\tilde E(\mathbf r)\, \tilde E(\mathbf r')}{(\tilde k^2 - k^2) \int \varepsilon\, \tilde E^2 \, d^2\mathbf r}.
\end{equation}
On resonance, \(k = k^*\), so
\begin{equation} \begin{aligned} \label{eq:detuning}
    \tilde k^2 - k^{*2} &= 2 i k^* k_i - k_i^2 \\
    &= -\frac{i k^{*2}}{Q} \left( 1 - \frac{i}{4Q} \right) \approx -\frac{i k^{*2}}{Q}.
\end{aligned} \end{equation}
Substituting~\eqref{eq:detuning} into~\eqref{eq:G_single_mode} at \(\mathbf r = \mathbf r' = 0\) and using~\eqref{eq:mode_vol} gives \(G(0, 0) \approx iQ/(k^{*2} V)\), so that \(\rho = (Q/k^{*2}) \Re(1/V)\).
In 2D, \(\rho_0 = \Im G_0(0, 0) = \tfrac14\), so
\begin{equation} \label{eq:purcell_2d}
    \frac{\rho}{\rho_0} = \frac{4Q}{k^{*2}}\, \Re \frac{1}{V}.
\end{equation}
Every \(\rho/\rho_0\) we report is the power extracted from the source at \(\omega^*\), divided by that extracted in vacuum at the design frequency \(\omega\) on the same grid.
The power extracted from a line current, \(-\tfrac12 \Re \int \mathbf J^* \cdot \mathbf E\), is proportional to \(\omega \Im G(0, 0)\), so this ratio is~\eqref{eq:purcell_2d} multiplied by \(\omega^*/\omega\), which lies within \(1\%\) of unity for every structure.
This calculation matches the formula above very closely for all structures.

\section{Symmetry Breaking in Lossy Rings} \label{asec:lossy}

\begin{figure}
    \centering
    \includegraphics[width=\linewidth]{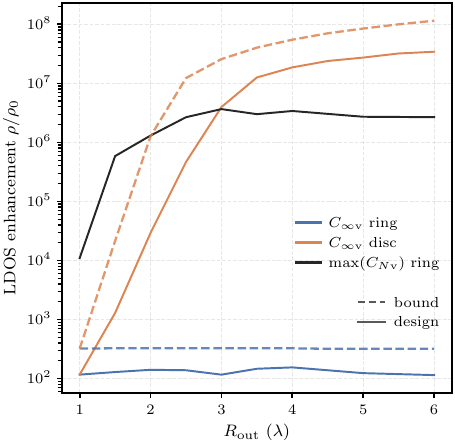}
    \caption{\textbf{Symmetry breaking in lossy rings.}
    LDOS enhancement \(\rho/\rho_0\) as a function of \(R_\textrm{out}\) for rings of width \(w = 0.8\lambda\) and susceptibility \(\chi = 5 + 10^{-6}i\), with \(\Cinfv\) discs of inner radius \(R_\textrm{in} = 0.2\lambda\) for reference.
    Dashed: \(\Cinfv\) limits; solid: inverse designs.
    The \(\max(\Cv{N})\) curve is the best ring over every order \(2 \leq N \leq 50\), all of which were computed at every \(R_\textrm{out}\).}
    \label{fig:lossy}
\end{figure}

Figure~\ref{fig:lossy} repeats the ring comparison of Fig.~2(a) for a lossy material, \(\chi = 5 + 10^{-6}i\), with the same source, inverse design procedure, and resonance criterion.
Limits are computed only for cylindrically symmetric (\(\Cinfv\)) rings and discs.
As in the lossless case, the \(\Cinfv\) ring limit is flat in \(R_\textrm{out}\), whereas symmetry-broken rings exceed it at every \(R_\textrm{out}\), by a factor of \(34\) at \(R_\textrm{out} = \lambda\) and by more than \(10^3\) from \(1.5\lambda\) on; symmetry breaking is therefore necessary for optimal LDOS enhancement in lossy rings.
For \(R_\textrm{out} \leq 1.5\lambda\) the symmetry-broken rings also exceed the \(\Cinfv\) limit for the \textit{disc} containing them, by factors of \(34\) at \(R_\textrm{out} = \lambda\) and \(27\) at \(1.5\lambda\).

\bibliographystyle{apsrev4-2}
\bibliography{refs}